\documentclass{pramana}

\usepackage{graphicx,amsmath,bm}
\usepackage{xcolor}

\begin{document}

\title{Quantifying the particle aspect of quantum systems}


\author{Sreetama Das\textsuperscript{1, *}, Indranil Chakrabarty\textsuperscript{2, 3}, Arun Kumar Pati\textsuperscript{4}, Aditi Sen (De)\textsuperscript{5} \and Ujjwal Sen\textsuperscript{5}}
\affilOne{\textsuperscript{1} Istituto Nazionale di Ottica del Consiglio nazionale delle Ricerche (CNR-INO), I-50019 Sesto Fiorentino, Italy\\}
\affilTwo{\textsuperscript{2} Centre for Quantum Science and Technology, International Institute of Information Technology Hyderabad, Gachibowli, Hyderabad 500032, Telangana, India.\\}
\affilThree{\textsuperscript{3} Center for Security, Theory and Algorithmic Research, International Institute of Information Technology Hyderabad, Gachibowli, Hyderabad 500032, Telangana, India\\}
\affilFour{\textsuperscript{4} Center for Quantum Engineering, Research and Education (CQuERE), TCG CREST, Salt lake, Kolkata 700090, India\\}
\affilFive{\textsuperscript{5} Harish-Chandra  Research  Institute, A CI of Homi Bhabha National Institute, Chhatnag Road, Jhunsi, Allahabad 211 019, India}


\twocolumn[{

\maketitle

\corres{sreetama.das21@gmail.com}


\begin{abstract}
The possibility of a quantum system to exhibit properties that are akin to both the classically held notions of being a particle and a wave, is one of the most intriguing aspects of the quantum description of nature. These aspects have been instrumental in understanding paradigmatic natural phenomena as well as to provide nonclassical applications. A conceptual foundation for the wave nature of a quantum state has recently been presented, through the notion of quantum coherence. We introduce here a parallel notion for the particle nature of a quantum state of an arbitrary physical system. We provide elements of a resource theory of particleness, and give a quantification of the same. Finally, we provide evidence for a complementarity between the particleness thus introduced, and the coherence of an arbitrary quantum state. 
\end{abstract}



}]



\section{Introduction}
The wave-particle duality is one of the core aspects of quantum mechanics. This tells us about a profound and, within classical intuitions, paradoxical behavior of nature by which a quantum system exhibits the property of being both a wave and a particle. 
The wave nature of a quantum entity is distinctly revealed in the double-slit experiment.
The wave aspect of a quantum system makes it pass through both slits at the same time, resulting in interference. This phenomenon can certainly be explained if the corresponding quantum system is considered to be a classical wave, but not if it was a classical particle.

On the other hand, the photoelectric effect is an example of a phenomenon where a quantum system exhibits particle-like characteristics. 
Contrary to the double-slit experiment, the photoelectric effect would not be explainable if we consider the quantum system as a classical wave. 
It is intriguing that there are quantum systems -- photons -- that exhibit both wave phenomenon via interference in double-slit experiment and particle aspect in the photoelectric effect. 

This duality of photons was known from the beginnings of quantum mechanics, and indeed was one of the main reasons for the latter's formulation. There has been tremendous achievements in this arena both on experimental and theoretical fronts. Let us, in particular, mention about the recent experiment that claim to  have ``coaxed'' an object of 10 kilograms to exhibit signature quantum properties~\cite{barafer-rasta}. Meanwhile, at least twelve photons have been simultaneously used to prepare an
entangled quantum state~\cite{dedar-charan-kabi-gaan-gaan-prante}.
There are, on the other hand, important theoretical developments that try to understand the meaning of macroscopicity within the quantum domain and try to find the boundary of the latter, if existing (see e.g.~\cite{pranter-kabi-kabe-asben-britye}).

The wave nature of a quantum system was observed 
long back,
and the quantum formalism was found to be able to 
incorporate it within its folds. However, a more careful conceptual foundation and quantification of the wave nature, that is independent of any particular  experiment, was presented only a few years back, 
where ``quantum coherence'' was quantified using a resource-theoretic framework
\cite{eibar-bachhadhan-chat-bagale, nijer-Dhak}.
%
%

It should be stressed that while  coherence can be viewed as a fundamental signature of nonclassicality in physical systems, the signature becomes substantial only if the \emph{same} system can also exhibit particle-like phenomena, as otherwise, the coherence phenomena can be explained by using a classical wave description of the system.

The particle nature of a quantum system, while again having been observed and incorporated into the quantum formalism since its beginnings, to the best of our knowledge, has not yet been conceptually formalized independent of the 
detailed aspects of the photoelectric effect. We hope to provide a way to bridge this gap, and provide elements of a resource theory of ``particleness'' of a quantum system. 
The motivation for the resource theory is taken from the 
photoelectric effect. As will become apparent, 
states having a non-zero value of ``particleness" defined by us,  act as a resource in the tasks where the photoelectric effect finds applications, e.g. the photoelectric diode, photomultiplier tube, photovoltaic cells, etc. These devices are used in a broad spectrum of 
electronic devices, 
solar cells, and in telecommunications.

Below we begin 
by providing a general model for detection of particle nature, 
providing the basis for conceptualizing the particle aspect of a quantum system, and the corresponding resource theory.
Although it is a toy model, it is independent of the details of the differences between the actual physical systems in which the phenomenon is observed, thus leading to wide applicability of the concepts provided.
We subsequently identify the ``free states'' of the resource theory, and follow this by 
considering the possible ``free operations''. 
Next, 
we consider the particular cases of two- and three-level systems respectively.
These discussions are given in the succeeding section (Sec.~\ref{prajapati}).
We then discuss about measures of particleness and about witnessing particleness (in Sec.~\ref{barkuchi}).
Moreover, we provide evidence for a complementarity relation between coherence and particleness for arbitrary quantum states, supported by 
numerical experiment using Haar-uniformly generated arbitrary three-dimensional pure and mixed quantum states.
The complementarity is discussed in Sec.~\ref{bishalakshmitala}.
We conclude with a discussion in Sec.~\ref{mallickpur}

%



\section{Model for conceptualizing particle aspect}
 \label{prajapati}
 Let us introduce here a toy model for detecting the particle aspect of a quantum system, taking inspiration from the photoelectric effect. 
The actual photoelectric interaction process is of course a much more complicated one and depends  on the actual materials involved. See e.g.,~\cite{shyam-thapa}. New aspects of the phenomenon keep being discovered (see e.g.~\cite{sharater-sashi}). A typical photoelectric interaction  involves an atom-photon interaction between the incoming photonic system and the material on which that impinges on. While the incoming photons can be modeled by a collection of quantum harmonic oscillators, and the material by a solid-state Hamiltonian, the interaction Hamiltonian can differ widely from one realization to another.

 The toy model used here is more in sync with the original formulation of the theory for the phenomenon, and this will help in quantifying the particle aspect of an \emph{arbitrary} quantum system, wherein a  \(d\)-level incoming quantum system impinges on  an effectively two-level ``solid state system''. The Hamiltonian of the incoming system is given by,
 \begin{equation}
 H=\sum _{n=0}^{d-1}\hbar \omega_n |n\rangle \langle n| ,
 \end{equation}
 with \(\{|n\rangle\}\) forming the energy eigenbasis and \(\hbar \omega_n\) representing the energy eigenvalues.
We consider the effective Hamiltonian of the solid state system to be \begin{equation}
H_{SS}=\hbar \omega|e\rangle \langle e|,
\end{equation}
where $ |e\rangle$ is the excited state of the effective two-level solid state system. \(\hbar \omega\) represents the energy gap between the two levels of the solid state system. In a more realistic situation, there can be a band of levels near the zero level energy of our effective solid state system, which is being approximated here by a single energy level with zero energy. Similarly, a possible band of energies near the excited state energy is being approximated here by a single excited state with energy \(\hbar \omega\). 
More generally, there can be metastable states between the two bands, and these are being not considered in this toy model. We do not explicitly write the interaction Hamiltonian. Instead, similar to what happens in case of the photoelectric effect, we assume that if there is an incoming state $\rho_{in}$ (on \(\mathbb{C}^d\)) such that  \begin{equation} \mbox{Tr}( \rho_{in}  H)> \hbar \omega,
\end{equation}
then the incoming state has a nonzero \emph{``particleness''}. 
In reality, the photocurrent also depends on the momentum of the involved electrons perpendicular to the surface of separation between the two systems involved, and this is ignored in the current formulation.
For simplicity, we  assume the ``zero detuning'' scenario, so that  $\omega_n=n\omega$ for $n=0,1, \ldots, d-1$.

Quantum coherence of a quantum state depends not only on the state but also on the basis. Indeed, the slits in an interference experiment defines such a basis, and the interference pattern changes depending on the 
character of the slits. 
Similarly, the particleness of a quantum state depends not only on the state but also on the Hamiltonians of the incoming system and the solid state. We have ignored the transfer mechanism of the energy from the impinging system to the solid state in the toy model. 




 \subsection{Free states} In any resource theory, 
an 
important aspect is to characterize the states which will not act as a resource -- 
the so-called \emph{``free states''}, which we denote as  $\rho_f$. If we consider particleness to be a resource of any given quantum system, the free states will be those states which cannot exhibit particleness of the system. This depends on the triplet consisting of 
\begin{itemize}
\item the state of the incoming system (\(\rho_{in}\)),
\item the Hamiltonian of the incoming system (\(H\)), and
\item the ``threshold energy '' (\(\hbar \omega\)) of the solid state system.
\end{itemize}

The free states will be 
those 
for which the energy content of the state 
is less than or equal to $ \hbar \omega$. Denoting the set of free states as \(F_S\), we have 
\begin{equation}F_S= \{\rho_f| \mbox{Tr}( \rho_f  H) \leq \hbar \omega \}.
\end{equation}
It is interesting to note that the set $F_S$ is a convex set and the corresponding ``\emph{edge states}'' 
are those for which the energy of the system is exactly equal to $ \hbar \omega$.  The definition is independent of the occupation number description of photonic modes, and so an incoming photonic mode having a 
non-zero particle number (occupation number), can be a free state according to our formalism, if the energy of the state is less than the band-gap of solid state material.


The states $|0\rangle$ and $|1\rangle$ are free states for any dimension, \(d\), of the incoming quantum system, and  $|1\rangle$ is an edge state  therein. 
For $n=2, \ldots, d$, the states $|n\rangle$ are resource states, i.e., have nonzero particleness. 
We remember that the states \(\{|n\rangle\}\) forms the eigenbasis of the Hamiltonian \(H\). 
If we consider the mixed state 
\begin{equation}
\rho^p_f=p|0\rangle \langle 0|+(1-p)|1\rangle \langle 1|
\end{equation}
($p \in [0,1]$), we find that  
\begin{equation}
\mbox{Tr} \left( \rho^p_f H\right)=\hbar \omega (1-p),
\end{equation}
so that \(\rho^p_f\) lies in the interior of 
the set of free states for all $d$, unless $p=0$.
%
Interestingly, the state  \begin{equation}
\frac{1}{3}I_3=\frac{1}{3}\left(|0\rangle \langle 0|+|1\rangle \langle 1|+|2\rangle \langle 2|\right)
\end{equation}
is an edge state for any input system, since \begin{equation}
\mbox{Tr} \left( \frac{1}{3}I_3 H\right)=\hbar \omega .
\end{equation}
A convex combination of this state and the state $|1\rangle$, i.e, a state of the form \begin{equation}
\frac{p}{3}I_3 +(1-p)1\rangle \langle 1|
\end{equation}
($p \in [0,1]$), is an edge state for any \(p\) and any input dimension, \(d\). The facts that \(\frac{1}{3}I_3\) and \(|1\rangle\) are edge states depends 
on our choice of equally spaced energy levels and the zero detuning. Changing the Hamiltonians will change the status of the free as well as edge states.
A schematic representation of the state space marking the free and resourceful states for a three-dimensional quantum system is presented in Fig.~1.

\subsection{Free operations} 

Along with free states, it is also important in any  resource theory to characterize the set of quantum operations which will keep free states as free states, and these operations are usually referred 
 as ``\emph{free operations}''. 

Let us now 
define our set of 
free operations in this resource theory as those collections of Kraus operators, $K_n$, for which \(\sum_n K_n^\dagger K_n\) is the identity on \(\mathbb{C}^d\) (``completeness condition''), and the energy of any free state is bounded above by  \(\hbar \omega\) even after the application of individual Kraus operators.
In other words, the set of free operations can be given by 
\begin{eqnarray}
F_O= \left\{\left\{K_n\right\}| (\mbox{Tr}(\sum_{\tilde{n}} K_{\tilde{n}} \rho_f  K_{\tilde{n}}^{\dagger} H) \leq \hbar \omega \;, ~\forall \rho_f \in F_S\right\},\nonumber\\
\end{eqnarray}
where the completeness condition is implicitly assumed. The summation over \(\tilde{n}\)
is over a subset, possibly proper, of elements of the entire set \(\{K_n\}\), and where a normalization factor \(\mbox{Tr}(\sum_{\tilde{n}} K_{\tilde{n}} \rho_f  K_{\tilde{n}}^{\dagger} )\) is assumed but kept silent in the notation.  
It is evident from the above definition 
that energy-invariant quantum operations form a class of free operations. Precisely, these are the Kraus operator sets for which 
\begin{equation}
\mbox{Tr}(\sum_{\tilde{n}} K_{\tilde{n}} \rho_f  K_{\tilde{n}}^{\dagger} H) = \mbox{Tr}(\rho_f H)\end{equation}
for all \(\rho_f \in F_S\).

A subset of the class of energy-invariant free operations are ones for which the Kraus operators commute with the Hamiltonian of the incoming system.



\subsection{Qubits}
If the incoming quantum system is a two-level system, let us write the corresponding state as 
$|\psi \rangle= a |0\rangle +b |1\rangle$ ($a$ and $b$ are the  amplitudes, $|a|^2 +|b|^2 =1$). In this case, the Hamiltonian operator determining the energy of the input is given by \begin{equation}
H=\hbar \omega |1\rangle \langle 1|.
\end{equation}
In this scenario, for all pure states $|\psi \rangle$, we have $\langle \psi |H|\psi \rangle= |b|^2\hbar \omega$, so 
that 
\begin{equation}
\langle \psi |H|\psi \rangle \leq \hbar \omega \end{equation}
(since $|b|^2 \leq 1$). This implies all the pure states are  free states, and the state $|1\rangle$ lies at the edge. Since mixed states are convex combinations of pure states, they will also lie in the set of free states, and since there is only a single pure edge state, there are no non-pure edge states. 

All two-level quantum systems are therefore devoid of any particleness. The scenario changes if we change the condition 
\(\omega_n = n \omega\) within which we are working. This condition is just an algebraic artefact that makes the discussion simpler, and changing it can easily result in qubit states having nonzero particleness.



\subsection{Qutrits}
Next we consider physical situation where a solid state system is exposed to a quantum source which is a three-level quantum system (e.g., a ladder-type three-level photonic system). This is the lowest dimension where we obtain 
quantum states having a nonzero particleness (under the assumption that the condition \(\omega_n=n\omega\) is still valid), and they appear with a finite volume. The  Hamiltonian 
of the quantum system is
given by   
\begin{equation}
H=\hbar \omega|1\rangle \langle 1|+2\hbar \omega |2\rangle \langle 2|.
\end{equation}
A pure state of the incoming system can be expressed as 
$|\psi \rangle= a |0\rangle +b |1\rangle+c |2\rangle $, where $a$, $b$, and $c$ are the amplitudes, with  $|a|^2 +|b|^2 +|c|^2 =1$.  
The solid state system on which the incoming particles are incident is still described by the Hamiltonian $H_{SS}=\hbar \omega|e\rangle \langle e| $.
There is no emission of particles (e.g., electrons)  from the zero-energy band of the solid state system unless the energy content of the incoming state is more than $\hbar\omega$, and hence there is no signature particle aspect of the system before that. It is easy to see that the pure state $|\psi \rangle$
is free if and only if 
\begin{equation}
|c| \leq |a|.
\end{equation}
For a general three-level  state \(\rho\), it is free if 
\begin{equation}
\rho_{11}
+ 2 \rho_{22}
\leq 1,
\end{equation}
where \(\rho_{nn}\) is the \(n\)\textsuperscript{th} diagonal element of \(\rho\), when written in the energy eigenbasis.
%


 \section{Measure and Detection of Particleness}
 \label{barkuchi}
 
 In the next step, we look for possible ways to quantify particleness 
of the incoming quantum system $\rho_{in}$, and for ways to witness the same.

One way to quantify the resource can be to use a concept akin to the definition of distillable entanglement \cite{boRda}. In that case, one begins by identifying a state that is the most resourceful, which in our case can be the highest eigenstate of the incoming system Hamiltonian. 
``Distillable particleness'' of an input state can then be defined as the asymptotic fraction of the most resourceful state, per input state, that can be obtained by free operations. In this paper, we follow a different track, viz. the distance-based approach. A similar approach is followed for defining the relative entropy of entanglement~\cite{kora-kagaz} and geometric 
measure of entanglement~\cite{rafi}.

Before moving to the explicit quantification of our measure, let us mention that concept of particles also appears in the de Broglie Bohm pilot wave hidden variable theory of quantum mechanics \cite{aj-bikeler-Daake}, where ``particles'' traverse the ``Bohmian'' trajectories, and their motion is guided, nonlocally, by a pilot wave.
Our notion of particleness, however, is conceptualized within quantum theory, without using a subquantum hidden variable theory.\\

\begin{figure}
\includegraphics[width=10cm,height=6cm,keepaspectratio]{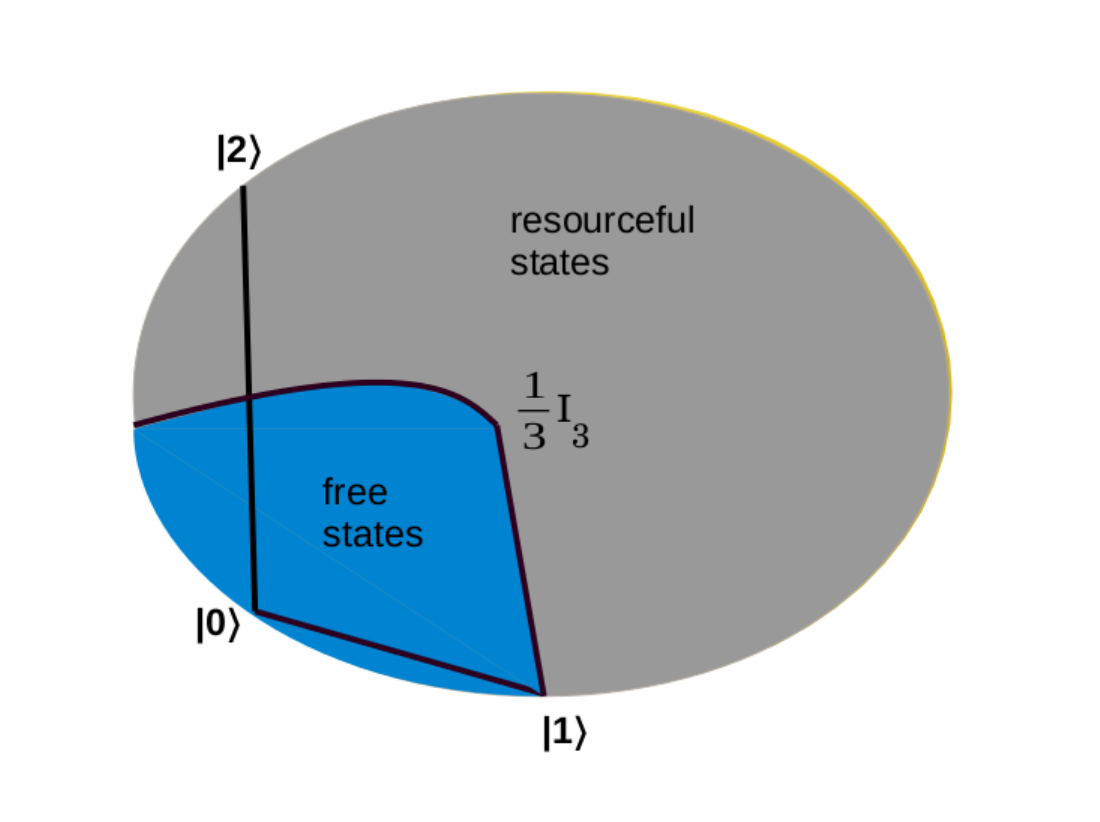} \label{tumi-ki-kebali}
\caption{Particleness of a qutrit: Free and resourceful states. We depict here the free and resourceful states in the space of density matrices of a three-level quantum system. Free states form a convex and compact set. The states \(\frac{1}{3}I_3\) and \(|1\rangle\) are edge states. The state \(|0\rangle\) is free but is not an edge state. The state \(|2\rangle\) is a resourceful state.}
\end{figure}

\subsection{Distance-based measure of particleness}  
We study here a distance-based measure of particleness,  $P_D(\rho_{in})$,  of a quantum state, \(\rho_{in}\) on \(\mathbb{C}^d\), 
given by  
\begin{equation}
P_D(\rho_{in})= \min_{\rho_f \in F_S} D(\rho_{in},\rho_f),
\end{equation}
where \(D(\rho,\sigma)\) is a distance function between arbitrary density matrices $\rho$ and $\sigma$.
Since we know that the state $\frac{1}{3}I_3=\frac{1}{3}(|0\rangle \langle 0|+|1\rangle \langle 1|+|2\rangle \langle 2|)$ is an edge state for any \(d\), we obtain the following result. \\


\noindent \textbf{Lemma:}  The 
particleness of an arbitrary quantum state \(\rho_{in}\)  is  bounded above by $D(\rho_{in},\frac{1}{3}I_3)$:
\begin{equation}
P_D\left(\rho_{in}\right) \leq D\left(\rho_{in},\frac{1}{3}I_3\right).
\end{equation}


Next we try to see that whether we can obtain a stronger 
bound, when we consider the distance of $\rho_{in}$ from a free state lying on the line joining $\rho_{in}$ with a free state in the interior of $F_S$. See Fig. 1 for a 
representation.
As noted before, the state $\rho_f^p= p|0\rangle \langle 0| +(1-p) |1\rangle \langle 1|$ is a free state for any \(p\) and any \(d\). 
And the state $|\psi \rangle= a |0\rangle +b |1\rangle+c |2\rangle $ ($a$, $b$, and $c$ are the amplitudes, with  $|a|^2 +|b|^2 +|c|^2 =1$), is free if and only if \(|c| \leq |a|\). Consider now the state
\begin{equation}
\rho (p,q)= q \rho_f^p + (1-q)|\psi \rangle \langle \psi| ,
\end{equation}
for \(q \in [0,1]\).
Considering a resourceful \(|\psi\rangle\), for any given \(p\), 
$\rho (p,q)$ is free if \begin{equation}
q\geq \frac{|c|^2-|a|^2}{p+|c|^2-|a|^2}.
\end{equation}
The equality sign  holds for a $\rho (p,q)$ on the edge. 
Therefore, the particleness  $P_D(|\psi \rangle)$ for a three-level pure state  $|\psi \rangle$, is bounded by 
\begin{equation}
\min_p D\left(|\psi \rangle \langle \psi |, \rho \left(p,q_p\right)\right),
\end{equation}
where 
\begin{equation}
q_p = \frac{|c|^2-|a|^2}{p+|c|^2-|a|^2}.
\end{equation}

\subsection{Detection of particleness and  existence of witness operators} 
The detection of resourceful states in any resource theory is of fundamental as well as of utilitarian significance. It is therefore interesting to find ways to detect resourceful states in the resource theory of particleness. 
A powerful and experimentally-friendly method to detect resourceful states in the resource theory of entanglement~\cite{aap-ki-akhon-mein} is by using the concept of entanglement witnesses, that may detect
entangled states prepared in a process if we have certain partial information about the process~\cite{aap-ki-akhon-mein, pinjar}. Below we show that one can similarly set up a witness for detecting particleness of a quantum state.\\

%


\noindent \textbf{Proposition:} $F_S= \{\rho_f| \mbox{Tr}( \rho_f  H) \leq \hbar \omega \}$ is convex and closed.\\

\noindent \textbf{Proof.} 
(i) To show that $F_S$ is convex, let us consider two free states, $\rho_1$ and $\rho_2$, of
$F_S$. Since  $\rho_1, \rho_2 \in F_S$, we have, $\mbox{Tr}( \rho_1  H) \leq \hbar \omega$ and $\mbox{Tr}( \rho_2  H) \leq \hbar \omega$.  
Let $\rho=\lambda \rho_1+(1-\lambda) \rho_2 $, where $0< \lambda <1$. But then, $\mbox{Tr}( \rho  H) = \mbox{Tr}( (\lambda \rho_1+(1-\lambda) \rho_2)  H) \leq \hbar \omega$, so that 
\(\rho \in F_S\).
\\ 

\noindent (ii) 
To prove the closedness of \(F_S\), 
we note that
\(\mbox{Tr}(\cdot H)\) is a continuous function of its argument. Therefore, the pre-image \(F_S\) of the closed set \((-\infty, \hbar \omega]\) is closed. \hfill \(\blacksquare\)\\

Since \(F_S\) is convex and closed, it is  possible to use the Hahn-Banach theorem \cite{nishiddhya-istehar} to provide the concept of witness operators for detecting states with nonzero particleness, similar to the concept of entanglement witnesses \cite{aap-ki-akhon-mein,pinjar}.


\begin{figure}
\includegraphics[width=11cm,height=6cm,keepaspectratio]{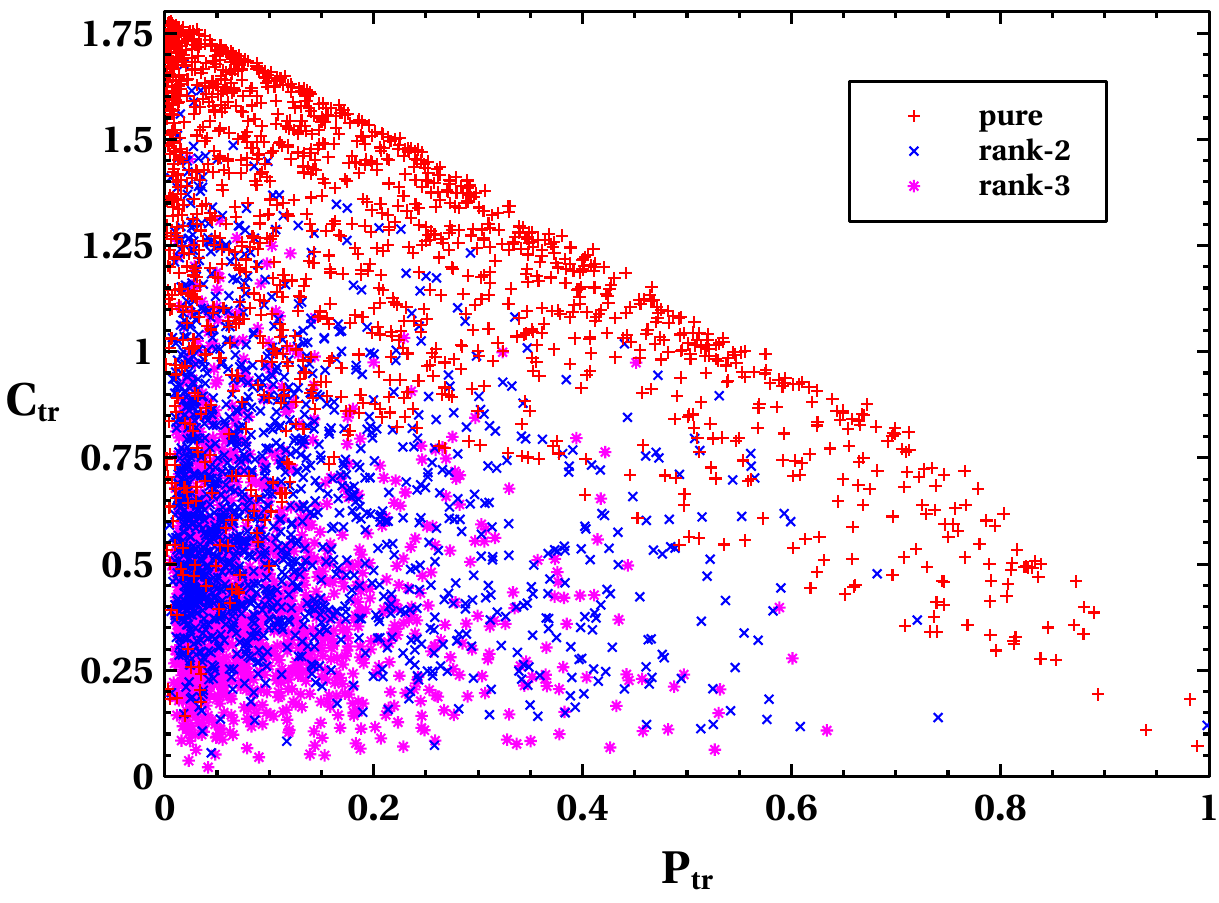} \label{complementarity}
\caption{Complementarity between particleness and quantum coherence. 
The points in the scatter diagram correspond to states of different ranks, and represents the coherence and particleness of those states. The measures for coherence and particleness used are the trace-norm coherence ($C_{tr}$) and trace-norm particleness ($P_{tr}$).
The horizontal axis represents particleness, while the vertical one represents quantum coherence. All quantities are dimensionless. For 
each rank, an order of  \(10^3\) states are generated Haar uniformly. We use red pluses for rank 1 states, blue crosses for 2,   and magenta asterisks for 3. For non-rank 1 states, the induced metric is used for the Haar-uniform generations \cite{zyc}.}
\end {figure}

\noindent 
\section{Complementarity between coherence and particleness} 
\label{bishalakshmitala}

The
wave-particle duality is an important aspect of experiments that formed
the very basis of the enunciation of the quantum theory of nature. And
while coherence has been 
regarded
as the measure of the wave nature of
quantum systems, we have argued that particleness is a measure of the
particle nature of the same. It is therefore conceivable that there will
appear a complementary relation between coherence and particleness for
arbitrary quantum states. In Fig. 2, 
we exhibit a 
(numerically) generated planar scatter diagram for Haar-uniformly generated arbitrary
three-dimensional quantum states that shows that such a complementarity
is indeed valid. We separately generate states of ranks 1, 2, and 3, Haar-uniformly in each case. 
The measures used to plot the diagram are respectively the trace-norm coherence ($C_{tr}$) and trace-norm particleness ($P_{tr}$), 
which are given by 
\begin{equation}
    A_{tr}(\rho) = \min_{\sigma \in \mathcal{F}_{S}} |\rho-\sigma|, \hspace{1cm} A=C,P, 
\end{equation}
and $|\rho-\sigma|=\text{Tr} \sqrt{(\rho-\sigma)^{\dagger}(\rho-\sigma)}$. Here, $\mathcal{F}_{S}$ is the set of free states of the corresponding measure, so that $\mathcal{F}_{S}=F_S$ for \(A=P\), and $\mathcal{F}_{S}$ is the set of incoherent states (i.e., the states diagonal in the basis \(\{|0\rangle, |1\rangle, |2\rangle\}\)) for \(A=C\) \cite{eibar-bachhadhan-chat-bagale, nijer-Dhak}. The numerically generated scatter diagram points lie below the line
\begin{equation}
P_{tr} + 1.3 C_{tr} \leq 1.8,
\end{equation}
where saturation is attained for certain pure states in \(\mathbb{C}^3\), with 
rank 2 states lying relatively away (and below) the bounding line in comparison to pure states, and rank 3 states being even farther away. 
This complementarity could, we believe, be yet another
face of the wave-particle duality of quantum physics \cite{gajanan, nijer-Dhak}. It should be noted that in a typical wave-particle duality relation, the ``waveness'' is posed against a path-distinguishability term. In an interference experiment,
almost all pure states have a certain amount of reduction in interference visibility and correspondingly a certain amount of which-path information. Our measure is however defined from the perspective of the photoelectric effect, where not all input photons can lead to a photocurrent.\\

\section{Conclusion}
\label{mallickpur}

The wave-particle duality is one of the crucial aspects of the edifice of quantum mechanics. While both the wave and particle aspects were well-known from the beginnings of quantum mechanics, their conceptual foundations were not formalized until recently, when the wave aspect was quantified, and called ``quantum coherence''. We have proposed 
a general framework for the particle aspect of an arbitrary quantum system, and 
provided components of 
a resource theory of particleness. In particular, we provided a depiction
of the qutrit space and the corresponding set of free states. 
We quantified the particleness of a state by using a distance-based measure, and also showed that the theory of entanglement witnesses can be carried over to the resource theory of particleness. 
We also indicated that there exists a  complementarity between 
the concepts of particleness 
and quantum coherence. 

The resource theory of particleness has marked similarities and differences with existing resource theories. In particular, unlike the resource theory of entanglement, the one of particleness does not only depend on the quantum state whose resource we are calculating. Rather, like for quantum coherence, in which the quantity for a given state depends on the ``preferred basis'', the particleness of a quantum state depends on the Hamiltonian of the incoming system (to which the quantum state belongs) and the Hamiltonian of the system on which it is impinging on.

Since the notions of quantum coherence and particleness depends on the choice of basis in the former and that of the incoming state and solid state Hamiltonians in the latter, one can associate a myriad of waves and particles for a single quantum. This is in sharp contrast to the notion of wave and particle that we have in the classical world.
We believe that our formalism  opens up a new avenue of exploration
of  ``particle'' nature and sets the stage for consideration of ``wave'' and ``particle'' natures
on an
equal footing.

\section*{Acknowledgement}
We acknowledge discussions with Nirman Ganguly, Chiranjib Mukhopadhyay, Rounak Mundra, and 
Kornikar Sen. The research of SD was supported in part by the INFOSYS scholarship for senior students. The authors of the Harish-Chandra Research Institute  acknowledge support from the Department of Science and Technology, Government of India through the QuEST grant.



\end{document}